\documentclass[12pt]{article}
\usepackage{amssymb}
\usepackage{hyperref}
\usepackage[pdftex]{graphicx}

\newcommand{\la}{\lambda}
\newcommand{\ga}{\gamma}

\newcommand{\dl}{\delta}
\newcommand{\bw}{{\bf w}}

\newcommand{\Tr}{{\rm Tr}}
\newcommand{\rmi}{{\rm i}}
\newcommand{\rmd}{{\rm d}}

\newtheorem{remark}{Remark}
\newtheorem{lemma}{Lemma}
\newtheorem{theorem}{Theorem}
\newtheorem{defin}{Definition}

\begin{document}
\title{Control landscapes for two-level open quantum systems}
\author{Alexander Pechen$^1$\thanks{E-mail: apechen@princeton.edu},
Dmitrii Prokhorenko$^2$, Rebing Wu$^1$\\ and Herschel
Rabitz$^1$\thanks{E-mail: hrabitz@princeton.edu}} \maketitle $^1$ Department
of Chemistry, Princeton University, Princeton, New Jersey 08544, USA

$^2$ Institute of Spectroscopy, Troitsk, Moscow Region 142190, Russia

\begin{abstract} A quantum control landscape is defined as
the physical objective as a function of the control variables. In this paper
the control landscapes for two-level open quantum systems, whose evolution is
described by general completely positive trace preserving maps (i.e., Kraus
maps), are investigated in details. The objective function, which is the
expectation value of a target system operator, is defined on the Stiefel
manifold representing the space of Kraus maps. Three practically important
properties of the objective function are found: (a) the absence of local
maxima or minima (i.e., false traps); (b) the existence of multi-dimensional
sub-manifolds of optimal solutions corresponding to the global maximum and
minimum; and (c) the connectivity of each level set. All of the critical
values and their associated critical sub-manifolds are explicitly found for
any initial system state. Away from the absolute extrema there are no local
maxima or minima, and only saddles may exist, whose number and the explicit
structure of the corresponding critical sub-manifolds are determined by the
initial system state. There are no saddles for pure initial states, one saddle
for a completely mixed initial state, and two saddles for partially mixed
initial states. In general, the landscape analysis of critical points and
optimal manifolds is relevant to explain the relative ease of obtaining good
optimal control outcomes in the laboratory, even in the presence of the
environment.
\end{abstract}

\section{Introduction}
A common goal in quantum control is to maximize the expectation value of a
given target operator by applying a suitable external action to the system.
Such an external action often can be realized by a tailored coherent control
field steering the system from the initial state to a target state, which
maximizes the expectation value of the target
operator~\cite{R,R0,R1,R2,RZ,R3,R4,R5,alessandro07}. Tailored coherent fields
allow for controlling Hamiltonian aspects (i.e., unitary dynamics) of the
system evolution. Another form of action on the system could be realized by
tailoring the environment (e.g., incoherent radiation, or a gas of electrons,
atoms, or molecules) to induce control through non-unitary system
dynamics~\cite{ice}. In this approach the control is the suitably optimized,
generally non-equilibrium and time dependent distribution function of the
environment; the optimization of the environment would itself be attained by
application of a proper external action. Combining such incoherent control by
the environment (ICE) with a tailored coherent control field provides a
general tool for manipulating both the Hamiltonian and dissipative aspects of
the system dynamics. A similar approach to incoherent control was also
suggested in~\cite{Romano} where, in difference with~\cite{ice}, finite-level
ancilla systems are used as the control environment. The initial state of the
field and the interaction Hamiltonian as the parameters for controlling
non-unitary dynamics was also suggested in~\cite{acim}. Non-unitary controlled
quantum dynamics can also be realized by using as an external action suitably
optimized quantum measurements which drive the system towards the desired
control goal~\cite{qm1,qm2,roa1,qm,feng}. General mathematical definitions for
the controlled Markov dynamics of quantum-mechanical systems are formulated
in~\cite{belavkin}.

In this paper we consider the most general physically allowed transformations
of states of quantum open systems, which are represented by completely
positive trace preserving maps (i.e., Kraus maps)~\cite{CPmap,AL,NiCh,breuer}.
A typical control problem in this framework is to find, for a given initial
state of the system, a Kraus map which transforms the initial state into the
state maximizing the expected value $\langle\Theta\rangle$ of a target
operator $\Theta$ of the system. Practical means to find such optimal Kraus
maps in the laboratory could employ various procedures such as adaptive
learning algorithms~\cite{R1,GA}, which are capable of finding an optimal
solution without detailed knowledge of the dynamics of the system. Kraus maps
can be represented by matrices satisfying an orthogonality constraint (see
Sec.~II), which can be naturally parameterized by points in a Stiefel
manifold~\cite{Stiefel}, and then various algorithms may be applied to perform
optimization over the Stiefel manifold (e.g., steepest descent, Newton
methods, etc. adapted for optimization over Stiefel
manifolds)~\cite{EdArSm,Manton02}.

The quantum control landscape is defined as the objective expectation value
$\langle\Theta\rangle$ as a function of the control variables. The efficiency
of various search algorithms (i.e., employed either directly in the laboratory
or in numerical simulations) for finding the minimum or maximum of a specific
objective function can depend on the existence and nature of the landscape
critical points. For example, the presence of many local minima or maxima
(i.e., false traps) could result in either permanent trapping of the search or
possibly dwelling for a long time in some of them (i.e., assuming that the
algorithm has the capability of extricating the search from a trap) thus
lowering the search efficiency. In such cases stopping of an algorithm at some
solution does not guarantee that this solution is a global optimum, as the
algorithm can end the search at a local maximum of the objective function.
{\it A priori} information about absence of local maxima could be very helpful
in such cases to guarantee that the search will be stopped only at a global
optimum solution. This situation makes important the investigation of the
critical points of the control landscapes. Also, in the laboratory, evidence
shows that it is relatively easy to find optimal solutions, even in the
presence of an environment. Explanation of this fact similarly can be related
with the structure of the control landscapes for open quantum systems.

The critical points of the landscapes for closed quantum systems controlled by
unitary evolution were investigated in~\cite{U1,U2,U3,U4,U5}, where it was
found that there are no sub-optimal local maxima or minima and only saddles
may exist in addition to the global maxima and minima. In particular, it was
found that for a two-level system prepared initially in a pure state the
landscape of the unitary control does not have critical points except for
global minima and maxima.

The capabilities of unitary control to maximize or minimize the expectation
value of the target operator in the case of mixed initial states are limited,
since unitary transformations can only connect states (i.e., density matrices)
with the same spectrum. In going beyond the latter limitations, the dynamics
may be extended to encompass non-unitary evolution by directing the controls
to include the set of Kraus maps (i.e., dual manipulation of the system and
the environment). Quantum systems which admit arbitrary Kraus map dynamics are
completely controllable, since for any pair of states there exists a Kraus map
which transforms one into the another~\cite{rong}.

In this paper the analysis of the landscape critical points is performed for
two-level quantum systems controlled by Kraus maps. It is found that the
objective function does not have sub-optimal local maxima or minima and only
saddles may exist. The number of different saddle values and the structure of
the corresponding critical sub-manifolds depend on the system initial state.
For pure initial states the landscape has no saddles; for a completely mixed
initial state the landscape has one saddle value; for other (i.e., partially
mixed) initial states the landscape has two saddle values. For each case we
explicitly find all critical sub-manifolds and critical values of the
objective as functions of the Stokes vector of the initial density matrix. An
investigation of the landscapes for multi-level open quantum systems with a
different method may also be performed~\cite{rw}. The absence of local minima
or maxima holds also in the general case although an explicit description of
the critical manifolds is difficult to provide for multi-level systems. The
absence of false traps practically implies the relative ease of obtaining good
optimal solutions using various search algorithms in the laboratory, even in
the presence of an environment.

It should be noted that the property of there being no false traps relies on
the assumption of the full controllability of the system, i.e., assuming that
an arbitrary Kraus map can be realized. Restrictions on the set of available
Kraus maps can result in the appearance of false traps thus creating
difficulties in the search for optimal solutions. Thus, it is important to
consider possible methods for engineering arbitrary Kraus type evolution of a
controlled system. One method is to put the system in contact with an ancilla
and implement, on the coupled system, specific unitary evolution whose form is
determined by the structure of the desired Kraus map~\cite{ref1} (see also
Sec.~II). Lloyd and Viola proposed another method of engineering arbitrary
Kraus maps, based on the combination of coherent control and
measurements~\cite{lloyd}. They show that the ability to perform a simple
single measurement on the system together with the ability to apply coherent
control to feedback the measurement results allows for enacting arbitrary
Kraus map evolution at a finite time.

A level set of the objective function is defined as the set of controls which
produce the same outcome value for $\langle\Theta\rangle$. We investigate
connectivity of the level sets of the objective functions for open quantum
systems and show that each level set is connected, including the one which
corresponds to the global maximum/minimum of the objective function.
Connectivity of a level set implies that any two solutions from the same level
set can be continuously mapped one into another via a pathway entirely passing
through this level set. The proof of the connectivity of the level sets is
based on a generalization of Morse theory. Experimental observations of level
sets for quantum control landscapes can be practically performed, as it was
recently demonstrated for control of nonresonant two-photon
excitations~\cite{roslund06}.

\begin{figure*}[t]\center
\includegraphics[scale=0.8]{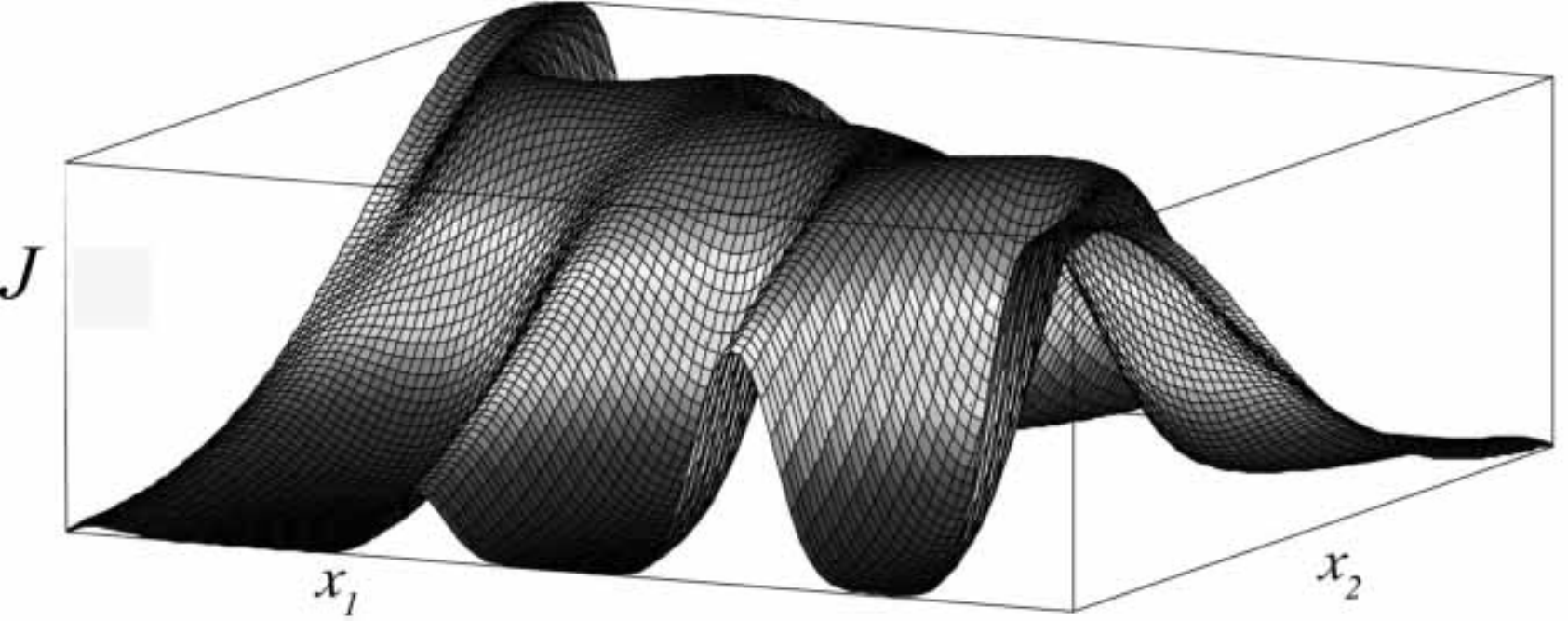}
\caption{This figure schematically illustrates the landscape $J$ as a function
of two controls $x_1$ and $x_2$. The figure shows the two main properties of
quantum-mechanical control landscapes for open quantum systems: (a) absence of
false traps and (b) connectivity of the sub-manifold of global maximum
solutions (a one dimensional curve at the top of the landscape in this
example).}\label{fig1}
\end{figure*}

In summary, the main properties of control landscapes for open quantum systems
are: (a) the absence of false traps; (b) the existence of multi-dimensional
sub-manifolds of global optimum solutions, and (c) the connectivity of each
level set. The proof of the properties (a)--(c) is provided in the next
sections for the two-level case. Figure~\ref{fig1} illustrates the properties
(a), (b), and connectivity of the manifold of global maximum solutions; the
figure does not serve to illustrate other properties such as connectivity of
each level set. It is evident that the function drawn on figure~\ref{fig1}
does not have local minima or maxima and the set of solutions for the global
maximum is a connected sub-manifold (a curve in this case). A simple
illustration is chosen for the figure since an exact objective function for an
$N$-level quantum system depends on $D=2N^4-2N^2$ real variables (such that
$D=24$ for $N=2$) and therefore can not be drawn.

The present analysis is performed in the kinematic picture which uses Kraus
maps to represent evolution of quantum open systems. An important future task
is to investigate the structure of the control landscape in the dynamical
picture, which can be based on the use of various dynamical master equations
to describe the dynamics of quantum open
systems~\cite{breuer,spohn,spohn2,dumcke,APV,ALV}. Such analysis may reveal
landscape properties for quantum open systems under (possibly, restricted)
control through manipulation by a specific type of the environment (e.g.,
incoherent radiation).

In addition to optimizing expected value of a target operator, a large class
of quantum control problems includes generation of a predefined unitary (e.g.,
phase or Hadamard)~\cite{NiCh} or a non-unitary~\cite{Tarasov} quantum gate
(i.e., a quantum operation). This class of control problems is important for
quantum computation and in this regard a numerical analysis of the problem of
optimal controlled generation of unitary quantum gates for two-level quantum
systems interacting with an environment is available~\cite{Grace1,Grace2}.

Although the assumption of complete positivity of the dynamics of open quantum
systems used in the present analysis is a generally accepted requirement, some
works consider dynamics of a more general form~\cite{nonCP,nonCP2}. Such more
general evolutions may result in different controllability and landscape
properties. For example, for a two-level open quantum system positive and
completely positive dynamics may have different accessibility
properties~\cite{romano05}. In this regard it would be interesting to
investigate if such different types of the dynamics have distinct essential
landscape properties.

In Sec.~\ref{sec1} the optimal control problem for a general $N$-level open
quantum system is formulated. Section~\ref{sec2} reduces the consideration to
the case of a two-level system. In Sec.~\ref{sec3} a complete description is
given of all critical points of the control landscape. The connectivity of the
level sets is investigated in Sec.~\ref{sec4}.

\section{Formulation for an $N$-level system}\label{sec1}
Let ${\cal M}_N$ be the linear space of $N\times N$ complex matrices. The
density matrix $\rho$ of an $N$-level quantum system is a positive component
in ${\cal M}_N$, $\rho\ge 0$, with unit trace, $\Tr\rho=1$ (Hermicity of
$\rho$ follows from its positivity). Physically allowed evolution
transformations of density matrices are given by completely positive trace
preserving maps (i.e., Kraus maps) in ${\cal M}_N$. A linear Kraus map
$\Phi:{\cal M}_N\to{\cal M}_N$ satisfies the following
conditions~\cite{CPmap}:
\begin{itemize}
\item{Complete positivity. Let $\mathbb I_n$ be the identity
matrix in ${\cal M}_n$. Complete positivity means that for
any integer $n\in\mathbb N$ the map $\Phi\otimes\mathbb
I_n$ acting in the space ${\cal M}_N\otimes {\cal M}_n$ is
positive.}

\item{Trace preserving: $\forall{\rho}\in{\cal M}_{N}$,
$\Tr\Phi(\rho)=\Tr\rho$.}
\end{itemize}
Any Kraus map $\Phi$ can be decomposed (non-uniquely) in the Kraus
form~\cite{choi,alessandro07}:
\begin{equation}\label{eq4}
\Phi(\rho)=\sum\limits_{l=1}^M K_l\rho K^\dagger_l,
\end{equation}
where the Kraus operators $K_l$ satisfy the relation
$\sum_{l=1}^M K^\dagger_lK_l=\mathbb I_N$. For an
$N$-level quantum system it is sufficient to consider at
most $M=N^2$ Kraus operators~\cite{choi}.

Let ${\cal H}_1=\mathbb C^N$ be the Hilbert space of the system under control.
An arbitrary Kraus map of the form~(\ref{eq4}) can be realized by coupling the
system to an ancilla system characterized by the Hilbert space ${\cal
H}_2=\mathbb C^M$, and generating a unitary evolution operator $U$ acting in
the Hilbert space of the total system ${\cal H}={\cal H}_1\otimes{\cal H}_2$
as follows~\cite{ref1}. Choose in ${\cal H}_2$ a unit vector $|0\rangle$ and
an orthonormal basis $|e_i\rangle$, $i=1,\dots,M$. For any $|\psi\rangle\in
{\cal H}_1$ let $U (|\psi\rangle\otimes|0\rangle)=\sum_{i=1}^M
K_i|\psi\rangle\otimes |e_i\rangle$. Such an operator can be extended to a
unitary operator in $\cal H$ and for any $\rho$ one has $\Phi(\rho)={\rm
Tr}_{{\cal H}_2}\,\{ U(\rho\otimes |0\rangle\langle 0|)U^\dagger\}$. Therefore
the ability to dynamically create, for example via coherent control, an
arbitrary unitary evolution of the system and ancilla allows for generating
arbitrary Kraus maps of the controlled system.

Let $\rho_0$ be the initial system density matrix. A typical optimization goal
in quantum control is to maximize the expectation value
$J=\langle\Theta\rangle$ of a target Hermitian operator $\Theta$ over an
admissible set of dynamical transformations of the system density matrices.
For coherent unitary control this expectation value becomes
\[
J[U]=\Tr[U\rho_0 U^\dagger\Theta]
\]
where $U=U(t,t_0)$ is a unitary matrix, $UU^\dagger=U^\dagger U=\mathbb I_N$,
which describes the evolution of the system during the control period from the
initial time $t_0$ until some final time $t$ and implicitly incorporates the
action of the coherent control field on the system.

In the present paper we consider general non-unitary controlled dynamics such
that the controls are Kraus maps, for which the parametrization by Kraus
operators is used. The corresponding objective function specifying the control
landscape has the form
\begin{equation}\label{eq8}
J[K_1,\dots,K_M]=\Tr\Bigl[\sum\limits_{l=1}^M K_l\rho_0
K^\dagger_l\Theta\Bigr]
\end{equation}
where the Kraus operators $\{K_l\}=\{K_l(t,t_0)\}$ describe evolution of the
open quantum system from an initial time $t_0$ until some final time $t$. The
control goal is to maximize the objective function over the set of all Kraus
operators $K_1,\dots,K_M$ satisfying $\sum_{l=1}^M K^\dagger_lK_l=\mathbb
I_N$, thereby forming a constrained optimization problem.

\begin{defin}
Let $\mathbb F$ be a field of real or complex numbers, i.e., $\mathbb
F=\mathbb R$ or $\mathbb F=\mathbb C$. A Stiefel manifold over $\mathbb F$,
denoted $V_k(\mathbb F^n)$, is the set of all orthonormal $k$-frames in
$\mathbb F^n$ (i.e., the set of ordered $k$-tuples of orthonormal vectors in
$\mathbb F^n$). The case $\mathbb F=\mathbb R$ (respectively, $\mathbb
F=\mathbb C$) corresponds to a real (complex) Stiefel manifold.
\end{defin}

Let $K$ be the $N\times (NM)$ matrix defined as $K=(K_1^{\rm T}\dots K_M^{\rm
T})$, where $K_l^{\rm T}$ is the transpose of matrix $K_l$ and $M$ is the
number of Kraus operators. Consider $N$ vectors $X_1,\dots, X_N\in\mathbb
C^{NM}$ with components $(X_i)_j=K_{ij}$, i.e., vector $X_i$ is the $i$-th row
of the matrix $K$. The constraint $\sum_{l=1}^M K^\dagger_lK_l=\mathbb I_N$ in
terms of the vectors $X_1,\dots, X_N$ takes the form $\langle
X_i,X_j\rangle=\dl_{ij}$, where $\dl_{ij}$ is the Kronecker delta symbol. This
constraint defines the complex Stiefel manifold $V_N(\mathbb C^{NM})$.
Therefore optimization of the objective function $J[K_1,\dots,K_M]$ defined by
Eq.~(\ref{eq8}) can be formulated as optimization over the complex Stiefel
manifold $V_N(\mathbb C^{NM})$.

\section{Two-level system}\label{sec2}
In the following we consider the case of a two-level system in detail. Any
density matrix of a two-level system can be represented as
\[
\rho=\frac{1}{2}[1+\langle{\bw,\sigma}\rangle]
\]
where
${\bf\sigma}=(\sigma_1,\sigma_2,\sigma_3)\equiv(\sigma_x,\sigma_y,\sigma_z)$
is the vector of Pauli matrices and ${\bf w}\in\mathbb R^3$ is the Stokes
vector, $\|{\bf w}\|\le 1$. Thus, the set of density matrices can be
identified with the unit ball in $\mathbb R^3$, which is known as the Bloch
sphere.

Any Kraus map $\Phi$ on ${\cal M}_2$ can be represented using at most four
Kraus operators
\[
K_l=\left(\begin{array}{cc}
x_{l1} & x_{l3}\\
x_{l2} & x_{l4}
\end{array}\right),\qquad l=1,2,3,4
\]
as $\Phi(\rho)=\sum_{l=1}^4K_l\rho K_l^\dagger$, where the Kraus operators
satisfy the constraint
\begin{equation}\label{eq1}
\sum\limits_{l=1}^4 K_l^\dagger K_l=\mathbb I_2
\end{equation}

Let $\rho_0$ be the initial system density matrix with Stokes vector $\bf
w=(\alpha,\beta,\gamma)$, where $\|\bw\|^2=\alpha^2+\beta^2+\gamma^2\le 1$,
and let $\Theta$ be a Hermitian target operator. The objective functional for
optimizing the expectation value of $\Theta$ has the form
$J[K_1,K_2,K_3,K_4;\rho_0,\Theta]=\sum_{l=1}^4\Tr [K_l\rho_0
K^\dagger_l\Theta]$. The control goal is to find all quadruples of Kraus
operators $(K_1,K_2,K_3,K_4)$ which maximize (or minimize, depending on the
control goal) the objective functional $J$. The goal of the landscape analysis
is to characterize all critical points of $J[K_1,K_2,K_3,K_4]$, including
local extrema, if they exist.

The analysis for an arbitrary $2\times2$ Hermitian matrix $\Theta$ can be
reduced to the case
\[
\Theta_0=\left(\begin{array}{cc}
1 & 0\\
0 & 0
\end{array}\right)
\]
which we will consider in the sequel. This point follows, as an arbitrary
Hermitian operator $\Theta\in{\cal M}_2$ has two eigenvalues $\la_1$ and
$\la_2$ and can be represented in the basis of its eigenvectors as
\[
\Theta=\left(\begin{array}{cc}
\la_1 & 0\\
0 & \la_2
\end{array}\right)
\]
where $\la_1\ge\la_2$. One has $\Theta=(\la_1-\la_2)\Theta_0+\la_2\mathbb I_2$
and
\begin{eqnarray*}
J[K_1,K_2,K_3,K_4;\rho_0,\Theta]&=&\sum\limits_{l=1}^4\Tr
[K_l\rho_0 K^\dagger_l\Theta]\\
&=&(\la_1-\la_2)\sum\limits_{l=1}^4\Tr [K_l\rho_0
K^\dagger_l\Theta_0]+\la_2\sum\limits_{l=1}^4\Tr
[K_l\rho_0
K^\dagger_l]\\
&=&(\la_1-\la_2)J[K_1,K_2,K_3,K_4;\rho_0,\Theta_0]+\la_2
\end{eqnarray*}
Therefore, the objective function for a general observable operator $\Theta$
depends linearly on the objective function defined for $\Theta_0$. We denote
$J[K_1,K_2,K_3,K_4;{\bf w}]:=J[K_1,K_2,K_3,K_4;\rho_0,\Theta_0]$. In the
trivial case $\Theta=\mathbb I_2$ the landscape is completely flat and no
further analysis is needed.

\section{The critical points of the objective function landscape}\label{sec3}
The Kraus operators for a two-level system can be parameterized by a pair of
vectors $X,Y\in\mathbb C^8=\mathbb C^4\oplus\mathbb C^4$ of the form
$X=u_1\oplus v_1$ and $Y=u_2\oplus v_2$, where
$u_1=(x_{11},x_{21},x_{31},x_{41})$, $v_1=(x_{12},x_{22},x_{32},x_{42})$,
$u_2=(x_{13},x_{23},x_{33},x_{43})$, and $v_2=(x_{14},x_{24},x_{34},x_{44})$.
The objective function in terms of these vectors has the form
\begin{equation}\label{eq7}
J[u_1,u_2,v_1,v_2;{\bf
w}]=\frac{1}{2}\Bigl[(1+\ga)\|u_1\|^2+(1-\ga)\|u_2\|^2+2{\rm
Re}[z_0\langle u_1,u_2\rangle]\Bigr]
\end{equation}
where $z_0=\alpha-\rmi\beta$, $\langle\cdot,\cdot\rangle$ and $\|\cdot\|$
denote the standard inner product and the norm in $\mathbb C^N$ (here the
numbers $\alpha,\beta,\gamma$ are the components of the Stokes vector
$\bw=(\alpha,\beta,\gamma)$ of the initial density matrix $\rho_0$, see
Sec.~\ref{sec2}). The constraint~(\ref{eq1}) in terms of the vectors $X$ and
$Y$ has the form $\|X\|=\|Y\|=1$, $\langle X,Y\rangle=0$ and determines the
Stiefel manifold ${\cal M}=V_2(\mathbb C^8)$. The matrix
constraint~(\ref{eq1}) in terms of the vectors $u_i$ and $v_i$ has the form
\begin{eqnarray}
\Phi_1(u_1,u_2,v_1,v_2)&:=&\|u_1\|^2+\|v_1\|^2-1=0\label{c1}\\
\Phi_2(u_1,u_2,v_1,v_2)&:=&\|u_2\|^2+\|v_2\|^2-1=0\label{c2}\\
\Phi_3(u_1,u_2,v_1,v_2)&:=&\langle u_1,u_2\rangle+\langle
v_1,v_2\rangle=0\label{c3}
\end{eqnarray}

If $z_0\ne 0$, then the objective function is diagonalized by introducing new
coordinates $(\tilde u_1,\tilde u_2,\tilde v_1,\tilde v_2)$ in $\mathbb
C^{16}$ according to the formulas
\begin{eqnarray} u_1&=&\mu \tilde u_1-\nu
\tilde u_2,\qquad
u_2=\frac{z_0^*}{|z_0|}\nu \tilde u_1+\frac{z_0^*}{|z_0|}\mu \tilde u_2\label{nc1}\\
v_1&=&\mu \tilde v_1-\nu \tilde v_2,\qquad
v_2=\frac{z_0^*}{|z_0|}\nu \tilde
v_1+\frac{z_0^*}{|z_0|}\mu \tilde v_2\label{nc2}
\end{eqnarray}
where $\mu=|z_0|/\sqrt{2\|\bw\|(\|\bw\|-\ga)}$ and
$\nu=|z_0|/\sqrt{2\|\bw\|(\|\bw\|+\ga)}$. The objective function in these
coordinates has the form
\begin{equation}\label{eq9}
J[x;{\bf w}]=\la_+\|\tilde u_1\|^2+\la_-\|\tilde u_2\|^2
\end{equation}
where $x=(\tilde u_1,\tilde u_2,\tilde v_1,\tilde v_2)\in{\cal M}$ and
$\la_\pm=(1\pm\|\bw\|)/2$. If $z_0=0$ and $\ga\ge 0$ (resp., $\ga<0$), then
the objective function~(\ref{eq7}) has the form~(\ref{eq9}) with $\tilde
u_i=u_i, \tilde v_i=v_i$ for $i=1,2$ (resp., $\tilde u_1=u_2,\tilde
u_2=u_1,\tilde v_1=v_2,\tilde v_2=v_1$). The
constraints~(\ref{c1})--(\ref{c3}) in the new coordinates have the same form
$\Phi_i(\tilde u_1,\tilde u_2,\tilde v_1,\tilde v_2)=0$ for $i=1,2,3$.

\begin{theorem}
Let ${\bf w}=(\alpha,\beta,\gamma)\in\mathbb R^3$ be a real vector such that
$\|{\bf w}\|\le 1$ and let $\la_\pm=(1\pm\|\bw\|)/2$. For any such $\bw$, the
global maximum and minimum values of the objective function $J[\tilde
u_1,\tilde u_2,\tilde v_1,\tilde v_2;{\bf w}]=\la_+\|\tilde
u_1\|^2+\la_-\|\tilde u_2\|^2$ are
\begin{eqnarray*}
\min\limits_{(\tilde u_1,\tilde u_2,\tilde v_1,\tilde
v_2)\in{\cal M}}J[\tilde u_1,\tilde u_2,\tilde v_1,\tilde
v_2;{\bf w}]&=&0\\
\max\limits_{(\tilde u_1,\tilde
u_2,\tilde v_1,\tilde v_2)\in{\cal M}}J[\tilde u_1,\tilde
u_2,\tilde v_1,\tilde v_2;{\bf w}]&=&1.
\end{eqnarray*}
The critical sub-manifolds and other critical values of $J$ in ${\cal M}$ are
the following:

{\bf Case 1.} $\bw=0$ (the completely mixed initial state). The global minimum
sub-manifold is ${\cal M}^{(0,0,0)}_{\rm min}=\{x\in{\cal M}\,|\,\tilde
u_1=\tilde u_2=0\}$. The global maximum sub-manifold is ${\cal
M}^{(0,0,0)}_{\rm max}=\{x\in{\cal M}\,|\,\tilde v_1=\tilde v_2=0\}$. The
objective function has one saddle value $J=1/2$ with the corresponding
critical sub-manifold ${\cal M}^{(0,0,0)}_{\rm saddle}=\{x\in{\cal M}\,|\,
\tilde u_2=z\tilde u_1,\, \tilde v_1=-z^*\tilde v_2,\, z\in\mathbb
C\}\bigcup\{x\in{\cal M}\,|\,\tilde u_1=\tilde v_2=0\}$. The Hessian of $J$ at
any point at ${\cal M}^{(0,0,0)}_{\rm saddle}$ has $\nu_+=6$ positive,
$\nu_-=6$ negative, and $\nu_0=16$ zero eigenvalues.

{\bf Case 2.} $0<\|\bw\|<1$ (a partially mixed initial state). The global
minimum sub-manifold is ${\cal M}^\bw_{\rm min}=\{x\in{\cal M}\,|\,\tilde
u_1=\tilde u_2=0\}$. The global maximum sub-manifold is ${\cal M}^\bw_{\rm
max}=\{x\in{\cal M}\,|\,\tilde v_1=\tilde v_2=0\}$. The objective function has
two saddle values:
\begin{equation}\label{s1}
J_\pm(\bw)=\frac{1\pm\|{\bf w}\|}{2}=\la_\pm.
\end{equation}
The corresponding critical sub-manifolds are ${\cal M}^{\bf w}_-=\{x\in{\cal
M}\,|\, \tilde u_1=\tilde v_2=0\}$ and ${\cal M}^{\bf w}_+=\{x\in{\cal M}\,|\,
\tilde u_2=\tilde v_1=0\}$. The Hessian of $J$ at any point at ${\cal
M}^\bw_-$ (resp., ${\cal M}^\bw_+$) has $\nu_+=8$ positive, $\nu_-=6$ negative
(resp., $\nu_+=6$ positive, $\nu_-=8$ negative), and $\nu_0=14$ zero
eigenvalues.

{\bf Case 3.} $\|\bw\|=1$ (a pure initial state). The global minimum
sub-manifold is ${\cal M}^\bw_{\rm min}=\{x\in{\cal M}\,|\,\tilde u_1=0\}$.
The global maximum sub-manifold is ${\cal M}^\bw_{\rm max}=\{x\in{\cal
M}\,|\,\tilde v_1=0\}$. The objective function has no saddles.

\end{theorem}
{\bf Proof.} The objective function has the form $J=\rho_{11}$, where
$\rho_{11}$ is the diagonal matrix element of the density matrix. Therefore
$0\le J\le1$ and the value $J=0$ (resp., $J=1$) corresponds to the global
minimum (resp., maximum).

The constraints can be included in the objective function~(\ref{eq9}) by
adding the term $\Phi[\tilde u,\tilde v,\eta]=\eta_1\Phi_1+\eta_2\Phi_2+2{\rm
Re}\,[\eta_3^*\Phi_3]$, where the two real and one complex Lagrange
multipliers $\eta_1,\eta_2$, and $\eta_3$ correspond to the two real and one
complex valued constraints $\Phi_1,\Phi_2$, and $\Phi_3$, respectively.
Critical points of the function $J$ on the manifold $\cal M$ are given by the
solutions of the following Euler-Lagrange equations for the functional
$\widetilde{J}[\tilde u,\tilde v,\la]=J[\tilde u,\tilde v]+\Phi[\tilde
u,\tilde v,\eta]$:
\begin{eqnarray} 0=\nabla_{\tilde u^*_1}
\widetilde{J}\Rightarrow\qquad 0&=&(\la_++\eta_1)\tilde u_1+\eta_3\tilde u_2\label{e1}\\
0=\nabla_{\tilde u^*_2}
\widetilde{J}\Rightarrow\qquad 0&=&\eta^*_3\tilde u_1+(\la_-+\eta_2)\tilde u_2\label{e2}\\
0=\nabla_{\tilde v^*_1}
\widetilde{J}\Rightarrow\qquad 0&=&\eta_1\tilde v_1+\eta_3\tilde v_2\label{e3}\\
0=\nabla_{\tilde v^*_2} \widetilde{J}\Rightarrow\qquad
0&=&\eta^*_3\tilde v_1+\eta_2\tilde v_2\label{e4}
\end{eqnarray}
where $\tilde u_1,\tilde u_2,\tilde v_1,\tilde v_2$ satisfy the
constraints~(\ref{c1})--(\ref{c3}). The proof of the theorem is based on the
straightforward solution of the system~(\ref{e1})--(\ref{e4}). The case~2 will
be considered first, followed by the cases~1 and~3.

{\bf Case 2.} $0<\|\bw\|<1$. Consider in $\cal M$ the open subset
$\mathcal{O}_1=\{x \in \mathcal{M}\;|\;\tilde v_1\ne 0,\;\tilde v_2\ne 0\}$.
Let us prove that the set of all critical points of $J$ in \(\mathcal{O}_1\)
is the set of all points of \(\mathcal{M}\) such that \(\tilde u_1=\tilde
u_2=0\).

Suppose that there are critical points in \(\mathcal{O}_1\) such that
\(\tilde{u}_1\neq 0\) or \(\tilde{u}_2\neq 0\). For such points the following
identity holds
\begin{equation}\label{3}
|\eta_3|^2=(\lambda_++\eta_1)(\lambda_-+\eta_2).
\end{equation}
In \(\mathcal{O}_1\), \(\tilde{v}_1\ne 0\) and therefore
\(|\eta_3|^2=\eta_1\eta_2\). This equality together with~(\ref{3}) gives
\begin{equation}\label{eq11}
\eta_2=-\la_-\left(1+\frac{\eta_1}{\la_+}\right)
\end{equation}
Suppose that \(\eta_3\neq 0\). Then, using~(\ref{e1})
and~(\ref{e3}), the constraint \({\Phi}_3\) gives
\[
(\lambda_++\eta_1)\|\tilde{u}_1\|^2+\eta_1\|\tilde{v}_1\|^2=0.
\]
Constraint \({\Phi}_1\) gives $\|\tilde{v}_1\|^2=1-\|\tilde{u}_1\|^2$, and
therefore $\eta_1=-\lambda_+\|\tilde{u}_1\|^2$. Similarly we find
$\eta_2=-\lambda_-\|\tilde{u}_2\|^2$. Substituting these expressions for
$\eta_1$ and $\eta_2$ into the (\ref{e1}) and (\ref{e2}) we find
\begin{equation}\label{eq2}
\left\{\begin{array}{c}
\sqrt{\lambda_-\lambda_+}\|\tilde{u}_1\|\|\tilde{u}_2\|^2=
\lambda_+(1-\|\tilde{u}_1\|^2)\|\tilde{u}_1\|\\
\sqrt{\lambda_-\lambda_+}\|\tilde{u}_1\|^2\|\tilde{u}_2\|=
\lambda_-(1-\|\tilde{u}_2\|^2)\|\tilde{u}_2\|
\end{array}\right.
\Rightarrow
\left\{\begin{array}{c}
\sqrt{\lambda_+}=\sqrt{\lambda_+}\|\tilde{u}_1\|^2+\sqrt{\lambda_-}\|\tilde{u}_2\|^2\\
\sqrt{\lambda_-}=\sqrt{\lambda_+}\|\tilde{u}_1\|^2+\sqrt{\lambda_-}\|\tilde{u}_2\|^2
\end{array}\right.
\end{equation}
This system of equations implies \(\lambda_-=\lambda_+\Leftrightarrow\bw=0\)
which is in contradiction with the assumption \(\|\mathbf{w}\|>0\) for the
present case. If \(\eta_3=0\), then it follows from~(\ref{e3}),~(\ref{e4})
that \(\eta_1=\eta_2=0\). In this case equations~(\ref{e1}) and~(\ref{e2})
have only the solution \(\tilde{u}_1=\tilde{u}_2=0\).

Points in ${\cal O}_1$ with $\tilde u_1=\tilde u_2=0$ form the global minimum
manifold \(\mathcal{M}^\bw_{\rm min}=V_2(\mathbb{C}^4)\), which is a Stiefel
manifold and hence is connected. In some small neighborhood of zero we can
choose \(\tilde{u}_1\) and \(\tilde{u}_2\) as normal coordinates. So
\(\mathcal{M}^\bw_{\rm min}\) is non degenerate. Similar treatment of the
region ${\cal O}_2=\{x\in{\cal M}\,|\,\tilde u_1\ne0,\,\tilde u_2\ne 0\}$
gives the global maximum manifold ${\cal M}^\bw_{\rm max}=\{x\in{\cal
M}\,|\,\tilde v_1=\tilde v_2=0\}$.

Now consider the region \(\mathcal{O}_3=\{x \in \mathcal{M}\,|\,\tilde u_2\neq
0,\,\tilde v_1\neq 0 \}\). In this region the objective function \(J\) has the
form
\begin{eqnarray}
J[\tilde{u}_1,\tilde{u}_2,\tilde{v}_1,\tilde{v}_2]=\lambda_-+
\lambda_+\|\tilde{u}_1\|^2-\lambda_-\|\tilde{v}_2\|^2.
\end{eqnarray}
Using the analysis for the region ${\cal O}_1$, we conclude that the objective
function has no critical points such that \(\tilde v_2\neq 0\) in ${\cal
O}_3$. Therefore all critical points in \(\mathcal{O}_3\) are in the
sub-manifold \(\mathcal{N}=\{x \in
\mathcal{M}\,|\,\tilde{v}_2=0\}\subset\mathcal{M}\). The restriction of \(J\)
to \(\mathcal{N}\) has the form
\begin{eqnarray*}
J[\tilde{u}_1,\tilde u_2,\tilde
v_1,\tilde{v}_2]|_{\mathcal{N}}=\lambda_-+\lambda_+\|\tilde{u}_1\|^2.
\end{eqnarray*}
Note that \(\mathcal{N}\) is a subset of all sets of
vectors \((\tilde{u}_1,\tilde{u}_2,\tilde{v}_1)\)
satisfying the constraints
\[
\|\tilde{u}_2\|^2=1,\qquad
\|\tilde{u}_1\|^2+\|\tilde{v}_1\|^2=1,\qquad
\langle\tilde{u}_1,\tilde{u}_2\rangle=0.
\]
It is clear from this representation of \(\mathcal{N}\) that \(\nabla
J|_{\mathcal{N}}=0\) if and only if \(\tilde{u}_1=0\). This gives the critical
sub-manifold ${\cal M}^\bw_-=\{x\in{\cal M}\,|\,\tilde u_1=\tilde v_2=0\}$.
The objective function has the value $J\mid_{{\cal M}^\bw_-}=\la_-$ on this
manifold.

To show that this is a saddle manifold, and not a local maximum or minimum, we
calculate the Morse indices of the objective function on ${\cal M}^\bw_-$ and
show that both positive and negative Morse indices are different from zero
(the Morse indices are the numbers of positive, negative and zero eigenvalues
of the Hessian of $J$ and positive and negative Morse indices determine the
number of local coordinates along which the function increases or decreases,
respectively). With regard to this goal, consider the manifold
\(\mathcal{K}:=\{x \in
\mathbb{C}^{16}\,|\,\Phi_1(\widetilde{u},\widetilde{v})=0,
\;\Phi_2(\widetilde{u},\widetilde{v})=0\}\). Let \(x \in \mathcal{M}\). Below
we introduce some coordinates in a neighborhood of \(x\) on \(\mathcal{K}\).

For any \(z \in \mathbb{C}^4\) such that \(z\neq0\) we define the unit vector
\(g(z)=z/\|z\|\in\mathbb{C}^4\). Let \(\varphi_i,\;i=1,\dots,7\) be some
coordinate system on \(S^7\) (embedded in \(\mathbb{C}^8\) as a unit sphere
with the origin at zero) in some neighborhood \(V_u\) of \(g(\tilde{u}_2(x))\)
and \(\psi_i,\;i=1,\dots,7\) be some coordinate system on \(S^7\) in some
neighborhood \(V_v\) of \(g(\tilde{v}_1(x))\). We will use the following
functions defined in some neighborhood of \(x\) on \(\mathcal{K}\) (\(z \in
\mathcal{K}\)):
\begin{eqnarray*}
\tilde{\varphi}_i(z)&=&\varphi_i\circ g\circ
\tilde{u}_2(z),\qquad i=1,\dots,7\\
\tilde{\psi}_i(z)&=&\psi_i\circ g\circ
\tilde{v}_1(z),\qquad i=1,\dots,7.
\end{eqnarray*}

Let $T_z S^7$ be the maximal complex subspace of the tangent space of $S^7$.
For each \(z \in V_u\) let \(x_1,\dots,x_6\) be coordinates on \(T_z {S^7}\)
and for each \(z \in V_v\) \(y_1,\dots,y_6\) be coordinates on \(T_z{S^7}\).

Let \(\tilde{x}_1,\dots,\tilde{x}_6\) and
\(\tilde{y}_1,\dots,\tilde{y}_6\) be functions on
\(\mathcal{K}\) defined as follows.

Let \(z=(\tilde{u}_1,\tilde{u}_2,\tilde{v}_1,\tilde{v}_2) \in \mathcal{K}\) be
in a small enough neighborhood of \(x\). By definition \(\rm Pr \mit_u\) is
the projection from \(\mathbb{C}^4\) to \(T_{g(\tilde{u}_2)}S^7\) and \(\rm Pr
\mit_v\) is the projection from \(\mathbb{C}^4\) to \(T_{g(\tilde{v}_1)}S^7\).
By definition
\begin{eqnarray*}
\tilde{x}_i&=&x_i\circ{\rm Pr}_u \circ \widetilde{u}_1, \qquad i=1,\dots,6,\\
\tilde{y}_i&=&y_i\circ{\rm Pr}_v \circ
\widetilde{v}_2,\qquad i=1,\dots,6.
\end{eqnarray*}
Now let \(\rm Pr \mit '_{u}\) and \(\rm Pr \mit '_{v}\) be the complex-valued
functions defined on \(\mathbb{C}^4\) by the formulas
\begin{eqnarray*}
{\rm Pr'}_{u}(f)&=&\langle
g(\widetilde{u}_2),f\rangle,\qquad f \in \mathbb{C}^4\\
{\rm Pr'}_{v}(f)&=&\langle
g(\widetilde{v}_1),f\rangle,\qquad f \in \mathbb{C}^4.
\end{eqnarray*}
By definition
\begin{eqnarray*}
p:={\rm Pr'}_{u}\circ\widetilde{u}_1,\qquad q:={\rm
Pr'}_{v}\circ\widetilde{v}_2.
\end{eqnarray*}
Thus, the functions
\(\tilde{\varphi}_i,\tilde{\psi}_i,\tilde{x}_k,\tilde{y}_l,p,q\), where
\(i,j=1,\dots,7\) and \(k,l=1,\dots,6\), are coordinates on \(\mathcal{K}\) in
some neighborhood of the point \(x\). Locally the manifold \(\mathcal{M}\) is
a sub-manifold of \(\mathcal{K}\) defined by the constraint \(\Phi_3=0\). In
our coordinates this constraint has a form
\[
p\left(1-\sum
\limits_{i=1}^{6}y_i^2-|q|^2\right)^{\frac{1}{2}}+
q\left(1-\sum
\limits_{i=1}^{6}x_i^2-|p|^2\right)^{\frac{1}{2}}=0.
\]
Therefore \(\tilde{\varphi}_i,\tilde{\psi}_i,\tilde{x}_k,\tilde{y}_l,p\),
where \(i,j=1,\dots,7\) and \(k,l=1,\dots,6\) are the coordinates on
\(\mathcal{M}\) in some neighborhood of \(x\). The second differential of
\(J\) at the point \(x\) in this coordinates has the form
\begin{eqnarray*}
\rmd^2J=\lambda_+\sum \limits_{i=1}^{6}\rmd x_i^2-
\lambda_- \sum \limits_{i=1}^{6}\rmd
y_i^2+(\lambda_+-\lambda_-)|\rmd p|^2.
\end{eqnarray*}
Since $\la_+-\la_-=\|\bw\|>0$ for the present case, the Morse indices of this
point are \(\nu_+=8,\;\nu_-=6\) (note that $p$ is a complex coordinate).

Similar treatment of the region ${\cal O}_4=\{x\in{\cal M}\,|\,\tilde u_1\ne
0,\,\tilde v_2\ne 0\}$ shows the existence of the critical sub-manifold ${\cal
M}^\bw_+=\{x\in{\cal M}\,|\,\tilde u_2=\tilde v_1=0\}$. This sub-manifold
corresponds to the critical value \(J\mid_{{\cal M}^\bw_+}=\lambda_+\) and its
Morse indices are \(\nu_+=6,\;\nu_-=8\). Since $\bigcup\limits_{i=0}^4{\cal
O}_i={\cal M}$, this concludes the proof for the case $0<\|\bw\|<1$.

{\bf Case 1.} \bw=0. Consider in $\cal M$ the open subset
$\mathcal{O}_1$.

Let $\eta_3=0$. Then in the region ${\cal O}_1$ Eqs.~(\ref{e3}) and~(\ref{e4})
imply that $\eta_1\tilde v_1=\eta_2\tilde v_2=0\Rightarrow\eta_1=\eta_2=0$.
Equations~(\ref{e1}) and~(\ref{e2}) for such $\eta_i$ have only the solution
$\tilde u_1=\tilde u_2=0$ which defines the global minimum manifold ${\cal
M}^{(0,0,0)}_{\rm min}=\{x\in{\cal M}\,|\,\tilde u_1=\tilde u_2=0\}$. Now let
$\eta_3\ne 0$ and $\tilde u_1\ne 0$ or $\tilde u_2\ne 0$. In this case
Eqs.~(\ref{e1})--(\ref{e4}) give $|\eta_3|^2=(1+\eta_1)(1+\eta_2)$ and
$|\eta_3|^2=\eta_1\eta_2$, which imply $\eta_2=-1-\eta_1$ and
$|\eta_3|^2=-\eta_1(1+\eta_1)$. Then Eqs.~(\ref{e1}) and~(\ref{e4}) have the
solution
\begin{equation}\label{eq3}
\tilde u_2=-\frac{1+\eta_1}{\eta_3}\tilde u_1=z\tilde
u_1,\qquad \tilde v_1=-\frac{\eta_2}{\eta^*_3}\tilde
v_2=-z^*\tilde v_2
\end{equation}
where we used the notation $z=-(1+\eta_1)/\eta_3\in\mathbb C/\{0\}$ and the
relation $-\eta_2/\eta^*_3=-z^*$. Note that for a given pair $(\tilde
u_1,\tilde v_2)\in\mathbb C^8$, $z$ can be any non-zero complex number such
that $(\tilde u_1,z\tilde u_1,-z^*\tilde v_2,\tilde v_2)\in{\cal M}$. The
solutions of the form~(\ref{eq3}) constitute the critical set ${\cal
T}=\{x\in{\cal O}_1\,|\,\tilde u_2=z\tilde u_1, \tilde v_1=-z^*\tilde v_2,
z\in\mathbb C\}\subset {\cal M}^{(0,0,0)}_{\rm saddle}$. A similar treatment
of the region ${\cal O}_2$ shows that the objective function in this region
has as critical points only the global maximum manifold ${\cal
M}^{(0,0,0)}_{\rm max}=\{x\in{\cal M}\,|\,\tilde v_1=\tilde v_2=0\}$ and the
set $\cal T$.

Now consider the region \(\mathcal{O}_3\).

Let $\eta_3=0$. Then in the region ${\cal O}_3$ Eqs.~(\ref{e2}) and~(\ref{e3})
imply $(1+\eta_2)\tilde u_2=\eta_1\tilde v_1=0\Rightarrow \eta_1=0$,
$\eta_2=-1$. The solution of Eqs.~(\ref{e1}) and~(\ref{e4}) for such values of
$\eta_i$ gives the critical set $\{x\in{\cal M}\,|\,\tilde u_1=\tilde
v_2=0\}\subset{\cal M}^{(0,0,0)}_{\rm saddle}$.

Let $\eta_3\ne 0$. The treatment is similar to the treatment of the case
$\eta_3\ne 0$ for the region ${\cal O}_1$ and gives the critical set $\cal T$.
A similar treatment of the region ${\cal O}_4$ shows that the set of critical
points of the objective function in this region is $\{x\in{\cal M}\,|\,\tilde
u_2=\tilde v_1=0\}\bigcup{\cal T}$.

Combining together the results for the regions ${\cal O}_1$, ${\cal O}_2$,
${\cal O}_3$, and ${\cal O}_4$, we find that the critical manifolds are the
global minimum manifold ${\cal M}^{(0,0,0)}_{\rm min}$, the global maximum
manifold ${\cal M}^{(0,0,0)}_{\rm max}$, and the set ${\cal
T}\bigcup\{x\in{\cal M}\,|\,\tilde u_2=\tilde v_1=0\}\bigcup\{x\in{\cal
M}\,|\,\tilde u_1=\tilde v_2=0\}\equiv{\cal M}^{(0,0,0)}_{\rm saddle}$. Since
$\bigcup\limits_{i=1}^4{\cal O}_i={\cal M}$, these manifolds are all critical
manifolds of the objective function $J$ for the case $\bw=0$. A simple
computation using the constraints~(\ref{c1})--(\ref{c3}) shows that the value
of the objective function at any point $x\in{\cal M}^{(0,0,0)}_{\rm saddle}$
equals to $1/2$, i.e., $J|_{{\cal M}^0}=1/2$.

Now we will find Morse indices of the critical manifold ${\cal
M}^{(0,0,0)}_{\rm saddle}$. An arbitrary point \(x=(u_1,u_2,v_1,v_2) \in
\mathcal{M}^{(0,0,0)}_{\rm saddle}\) can be moved into the point
\(\tilde{x}=(\tilde{u}_1,\tilde{u}_2,\tilde{v}_1,\tilde{v}_2) \in
\mathcal{M}^{(0,0,0)}_{\rm saddle}\) with \(\tilde{u}_1=0\), \(\tilde{v}_2=0\)
by the following transformation:
\begin{eqnarray*}
\tilde{u}_1&=&\alpha u_1+\beta u_2,\qquad
\tilde{u}_2=- \beta^*u_1+\alpha^* u_2,\\
\tilde{v}_1&=&\alpha v_1+\beta v_2,\qquad \tilde{v}_2=-
\beta^*v_1+\alpha^* v_2,
\end{eqnarray*}
where \(\alpha,\beta \in \mathbb{C}\),
\(|\alpha|^2+|\beta|^2=1\). For example, \(\alpha=-\beta
z\) for \(x=(u_1,zu_1,-z^*v_2,v_2) \in \mathcal{T}\).

As in the analysis of the Morse indices for the case~2, in some neighborhood
of \(\tilde{x}\) we can introduce the coordinates
\(\tilde{\varphi}_i,\tilde{\psi}_i, \tilde{x}_k,\tilde{y}_l,p, q\), where
\(i,j=1,\dots,7\) and \(k,l=1,\dots,6\). These coordinates satisfy the
constraint:
\[
p\left(1-\sum
\limits_{i=1}^6\tilde{y}_i^2-|q|^2\right)^{\frac{1}{2}}+q\left(1-\sum
\limits_{i=1}^6\tilde{x}_i^2-|p|^2\right)^{\frac{1}{2}}=0.
\]
The second differential of \(J\) in these coordinates has the form:
\begin{eqnarray}
{\rm d}^2J=\sum \limits_{i=1}^6 \rmd\tilde{x}_i^2-\sum
\limits_{i=1}^6 \rmd\tilde{y}_i^2+0\cdot |\rmd p|^2.
\end{eqnarray}
It is easy to see that the tangent space to \(\mathcal{M}^{(0,0,0)}_{\rm
saddle}\) at the point \(\tilde{x}\) is spanned by the vectors
\[
\frac{\partial}{\partial \tilde{\varphi}_i},\; \frac{\partial}{\partial
\tilde{\psi}_i},\; \frac{\partial}{\partial \rm Re \mit
p},\;\frac{\partial}{\partial \rm Im \mit p}.
\]
Therefore \(\mathcal{M}^{(0,0,0)}_{\rm saddle}\) is nondegenerate, \({\rm
dim}\,{\cal M}^{(0,0,0)}_{\rm saddle}=16\) and the Morse indices of
\(\mathcal{M}^{(0,0,0)}_{\rm saddle}\) are \(\nu_+=\nu_-=6\).

{\bf Case 3.} \( \|\bw\|=1\). In this case $\la_-=0$,
$\la_+=1$, and
\begin{eqnarray}
J[\tilde{u}_1,\tilde u_2,\tilde
v_1,\tilde{v}_2]=\|\tilde{u}_1\|^2.
\end{eqnarray}
Let \(\mathcal{U}_1=\{x \in \mathcal{M}\,|\,\tilde{v}_1\neq 0\}\). Clearly,
points in ${\cal U}_1$ with $\tilde u_1=0$ form the global minimum of the
objective. Assume that there are critical points in $\mathcal{U}_1$ such that
$\tilde u_1\ne 0$. For such points Eqs.~(\ref{e1})--(\ref{e4}) imply the
system of equations
\begin{eqnarray}
|\eta_3|^2&=&\eta_1\eta_2\\
|\eta_3|^2&=&\eta_2(1+\eta_1)
\end{eqnarray}
which has only the solutions with \(\eta_2=\eta_3=0\). But in the region
\(\mathcal{U}_1\), \(\tilde{v}_1\neq 0\) and therefore Eq.~(\ref{e3}) implies
\(\eta_1=0\). Then, Eq.~(\ref{e1}) for \(\eta_1=\eta_2=\eta_3=0\) has the
solution \(\tilde{u}_1=0\) which contradicts the assumption $\tilde u_1\ne 0$.
As a result, the only critical points in \(\mathcal{U}_1\) are with
\(\tilde{u}_1=0\). These points form the global minimum manifold ${\cal
M}^\bw_{\rm min}=\{x\in{\cal M}\,|\,\tilde u_1=0\}$. This manifold is
diffeomorphic to the space bundle with \(S^7\) as a base and \(S^{14}\) as a
fibre. Thus, ${\cal M}^\bw_{\rm min}$ is connected. We can use \(\tilde{u}_1\)
as normal coordinates in some neighborhood of ${\cal M}^\bw_{\rm min}$. Thus
${\cal M}^\bw_{\rm min}$ is nondegenerate.

The treatment of the region \(\mathcal{U}_2=\{x \in
\mathcal{M}\,|\,\tilde{u}_1\neq 0\}\) is equivalent to the previous
consideration. The critical points in this region form the global maximum
manifold ${\cal M}^\bw_{\rm max}=\{x\in{\cal M}\,|\,\tilde v_1=0\}$. Note that
\(\mathcal{U}_1\cup\mathcal{U}_2=\mathcal{M}\). Therefore, all critical points
of \(J\) correspond to the global minimum \(J=0\) and global maximum \(J=1\).
The critical manifolds corresponding to the minimum and the maximum are
connected and nondegenerate. $\Box$
\begin{remark}{\rm
The critical manifolds in terms of the original parametrization of the Kraus
operators by $(u_1,u_2,v_1,v_2)$ can be obtained by expressing $\tilde u_i$
and $\tilde v_i$ in terms of $u_i$ and $v_i$. If $z_0\ne 0$, then it follows
from~(\ref{nc1}) and~(\ref{nc2}) that
\begin{eqnarray*}
\tilde u_1&=&\mu u_1+\frac{z_0}{|z_0|}\nu u_2,\qquad
\tilde u_2=-\nu u_1+\frac{z_0}{|z_0|}\mu u_2\\
\tilde v_1&=&\mu v_1+\frac{z_0}{|z_0|}\nu v_2,\qquad
\tilde v_2=-\nu v_1+\frac{z_0}{|z_0|}\mu v_2
\end{eqnarray*}
Thus, for $z_0\ne 0$ and $0<\|\bw\|<1$ the critical manifolds are the
following: the global minimum ${\cal M}^\bw_{\rm min}=\{x\in{\cal
M}\,|\,u_1=u_2=0\}$, the global maximum ${\cal M}^\bw_{\rm max}=\{x\in{\cal
M}\,|\,v_1=v_2=0\}$, and the saddles ${\cal M}^\bw_\pm=\{x\in{\cal
M}\,|\,u_2=z_\pm u_1, v_1=-z^*_\pm v_2\}$. Here $z_\pm=z_0^*/(\ga\pm\|\bw\|)$.
For $z_0\ne 0$ and $\|\bw\|=1$ (hence $\ga\ne 1$), the critical manifolds are
${\cal M}^\bw_{\rm min}=\{x\in{\cal M}\,|\,u_2=z_0^*u_1/(\ga-1)\}$, ${\cal
M}^\bw_{\rm max}=\{x\in{\cal M}\,|\,v_2=z_0^*v_1/(\ga-1)\}$, and there are no
saddles.

If $z_0=0$ and $\ga\ge 0$, then $\tilde u_1=u_1$, $\tilde u_2=u_2$, $\tilde
v_1=v_1$, and $\tilde v_2=v_2$. Thus for $\ga=0$, ${\cal M}^{(0,0,0)}_{\rm
min}=\{x\in{\cal M}\,|\, u_1=u_2=0\}$, ${\cal M}^{(0,0,0)}_{\rm
max}=\{x\in{\cal M}\,|\, v_1=v_2=0\}$, and ${\cal M}^{(0,0,0)}_{\rm
saddle}=\{x\in{\cal M}\,|\, u_2=zu_1, v_1=-z^*v_2, z\in\mathbb
C\}\bigcup\{x\in{\cal M}\,|\, u_1=v_2=0\}$. For $0<\ga<1$ the critical
manifolds are ${\cal M}^{(0,0,\ga)}_{\rm min}=\{x\in{\cal M}\,|\,
u_1=u_2=0\}$, ${\cal M}^{(0,0,\ga)}_{\rm max}=\{x\in{\cal M}\,|\,
v_1=v_2=0\}$, and the saddles ${\cal M}^{(0,0,\ga)}_-=\{x\in{\cal M}\,|\,
u_1=v_2=0\}$ and ${\cal M}^{(0,0,\ga)}_+=\{x\in{\cal M}\,|\, u_2=v_1=0\}$. For
$\ga=1$, ${\cal M}^{(0,0,1)}_{\rm min}=\{x\in{\cal M}\,|\, u_1=0\}$ and ${\cal
M}^{(0,0,1)}_{\rm max}=\{x\in{\cal M}\,|\, v_1=0\}$.

If $z_0=0$ and $\ga<0$, then $\tilde u_1=u_2$, $\tilde u_2=u_1$, $\tilde
v_1=v_2$, and $\tilde v_2=v_1$. In this case for $-1<\ga<0$ the critical
manifolds are the following: ${\cal M}^{(0,0,\ga)}_{\rm min}=\{x\in{\cal
M}\,|\, u_1=u_2=0\}$, ${\cal M}^{(0,0,\ga)}_{\rm max}=\{x\in{\cal M}\,|\,
v_1=v_2=0\}$, ${\cal M}^{(0,0,\ga)}_-=\{x\in{\cal M}\,|\, u_2=v_1=0\}$ and
${\cal M}^{(0,0,\ga)}_+=\{x\in{\cal M}\,|\, u_1=v_2=0\}$. For $\ga=-1$, ${\cal
M}^{(0,0,-1)}_{\rm min}=\{x\in{\cal M}\,|\, u_2=0\}$ and ${\cal
M}^{(0,0,-1)}_{\rm max}=\{x\in{\cal M}\,|\, v_2=0\}$.}
\end{remark}
\begin{remark}{\rm
The values of the objective function at the saddle points satisfy the equality
$J_+(\bw)+J_-(\bw)=1$. This fact is a consequence of the more general symmetry
of the objective function, defined by the duality map $T:{\cal M}\to{\cal M}$
such that $T(u_1,u_2,v_1,v_2)=(v_1,v_2,u_1,u_2)$ as $J[x;\bw]+J[T(x);\bw]=1$
for any $x\in{\cal M}$. Thus, if the level set
$\Gamma_\bw(\alpha):=\{x\in{\cal M}\,|\,J[x,\bw]=\alpha\}$ for some value
$\alpha\in[0,1]$ is known then one immediately gets the level set for the
value $1-\alpha$ as $\Gamma_\bw(1-\alpha)=T(\Gamma_\bw(\alpha))$.}
\end{remark}

\section{Connectivity of the level sets}\label{sec4}
The level set $\Gamma_\bw(\mu)$ for an admissible objective value $\mu\in
[0,1]$ is defined as the set of all controls $x=(u_1,u_2,v_1,v_2)\in{\cal M}$
which produce the same outcome value $\mu$ for the objective function
$J[u_1,u_2,v_1,v_2;\bw]$, i.e., $\Gamma_\bw(\mu)=\{x\in{\cal
M}\,|\,J[x;\bw]=\mu\}$ (we omit the subscript $\bw$ in the sequel). In this
section it is shown that each level set for the function $J[\cdot;\bw]$ is
connected. This means that any pair of solutions in a level set $\Gamma(\mu)$
is connected via a continuous pathway of solutions entirely passing through
$\Gamma(\mu)$. Practically, connectivity of the level sets implies the
possibility to experimentally locate more desirable solutions via continuous
variations of the control parameters while maintaining the same value of the
objective function. The proof of the connectivity of the level sets for the
objective functions defined by~(\ref{eq7}) is based on generalized Morse
theory, which is presented in the remainder of this section. Theorem~\ref{t1}
below formulates the conditions for a generalized Morse function to have
connected level sets. These conditions are satisfied for the objective
function $J[\cdot,\bw]$ defined by~(\ref{eq7}), as stated in the end of this
section. Formulation of Theorem~\ref{t1} includes a very general class of
functions and can be applied to the investigation of connectivity of the level
sets for situations beyond the scope of this paper, including landscapes for
multilevel closed and open systems.

\subsection{Connectivity of level sets of generalized Morse
functions} Let \(M\) be a smooth compact manifold of dimension \(d\), and let
\(f\) be a smooth function \(f:M\rightarrow\mathbb{R}\). We suppose that the
critical set of \(f\), \(S:=\{x \in M| \rmd f(x)=0\}\) is a disjoint union of
smooth connected sub-manifolds \(C_i\) \((i=1,2,\dots,n)\) of dimension
\(d_i\). Let \(\mu_i=f|_{C_i}\).

For each point \(x \in C_i\) there exists an open neighborhood \(U\) of \(x\)
and a coordinate system \(\{x_l\}\) in \(U\) such that
\begin{eqnarray}
C_i\cap U=\{x \in U|x_{d_i+1}=\cdots=x_n=0\}.
\end{eqnarray}
Consider the following matrix
\begin{eqnarray}
J_i(x):= \left \| \frac{\partial^2 f(x)}{\partial
x_l\partial x_m} \right\|_{l,m=d_i+1,\dots,d}, \qquad x\in
C_i.
\end{eqnarray}
It is easy to see that if \(\{y_l\}\) is another coordinate system in \(U\)
such that
\begin{eqnarray}
C_i\cap U=\{y \in U|y_{d_i+1}=\cdots=y_n=0\},
\end{eqnarray}
and
\begin{eqnarray}
\widetilde{J}_i(x):= \left \| \frac{\partial^2
f(x)}{\partial y_l\partial y_m}
\right\|_{l,m=d_i+1,\dots,d}, \qquad x\in C_i
\end{eqnarray}
then
\begin{eqnarray}
\rm rank \mit J_i(x)=\rm rank \mit \widetilde{J}_i(x).
\end{eqnarray}
Therefore we can give the following

\begin{defin}The point \(x \in C_i\) is said to be
nondegenerate if \({\rm det}\, J_i(x)\neq 0\).
\end{defin}

\begin{defin} A critical sub-manifold \(C_i\) is said to be
nondegenerate if \(\forall x \in C_i\), \(x\) is a
nondegenerate point.
\end{defin}

Let \(x \in C_i\) and \(\lambda_i^+(x)\), \(\lambda_i^-(x)\) be the numbers of
positive and negative eigenvalues of the matrix \(J_i(x)\). It is clear that
\(\lambda_i^+(x),\lambda_i^-(x)\) do not depend on the choice of coordinate
system \(\{x_i\}\) in the neighborhood of \(x\). One can prove that
\(\lambda_i^+(x)\) and \(\lambda_i^-(x)\) do not depend on the point \(x \in
C_i\) (\(\lambda_i^+(x)\) and \(\lambda_i^-(x)\) are continuous and \(C_i\) is
connected.). Let \(\lambda_i^+:=\lambda_i^+(x)\) and
\(\lambda_i^-:=\lambda_i^-(x)\).
 \(\lambda_i^+\) and \(\lambda_i^-\) are called the indices of
 \(C_i\).

\begin{defin}
Let \(M\) be a smooth compact connected manifold and
\(f:M\rightarrow\mathbb{R}\). Suppose that the critical set of \(f\) is a
disjoint union of (compact) connected nondegenerate sub-manifolds \(C_i\). In
this case we say that \(f\) is a generalized Morse function. Sub-manifolds
\(C_i\) are called the critical sub-manifolds of \(f\).
\end{defin}

\begin{theorem}\label{t1} Let \(M\) be a smooth compact connected
manifold and \(f\) be a generalized Morse function. Let
\(C_i\), \(i=1,\dots,n\) be critical sub-manifolds of
\(f\) and \(\mu_i=f|_{C_i}\). We can assume that
\(\mu_{\min}:=\mu_1\leq\mu_2\leq\dots\leq
\mu_n=:\mu_{\max}\). Suppose that the sub-manifold
\(C_{\max}:=f^{-1}(\mu_{\max})\) is connected. Suppose
also that \(\forall i=1,\dots,n-1\) the indices
\(\lambda_i^+\geq2\), \(\lambda_i^-\geq 2\). Then
\(\forall \mu: \mu_{\min}\leq \mu\leq \mu_{\max}\) the set
\(\Gamma(\mu):=f^{-1}(\mu)\) is connected.
\end{theorem}
{\bf Proof.} We decompose the proof of the theorem into a sequence of several
Lemmas.

\begin{lemma}\label{m:1}
There exists an open neighborhood \(U\) of \(C_{\max}\) such that \(U\) is
diffeomorphic to some bundle \(\mathcal{E}\) with the base \(C_{\max}\) and
the fibre \(B_{d-d_n}\). Here \(B_k\) is a \(k\)-dimensional ball.
\end{lemma}
\textbf{Proof.} M is a compact. Therefore there exists a
Riemann metric \(g \in \rm sym (\mit T^\ast M\otimes
T^\ast M) \). (Here \(T^\ast M\) is a cotangent bundle of
\(M\).) By definition,
 \(\mathcal{L}\) is a restriction of the tangent bundle \(TM\) to
\(C_{\max}\). Let \(\mathcal{N}\) be a sub-bundle of \(\mathcal{L}\) such that
\(\forall x \in C_{\max}\) the fiber \(\mathcal{N}_x\) of \(\mathcal{N}\) over
\(x\) is a subspace of \(T_xM\) consisting of all vectors orthogonal to
\(T_xC_{\max}\). Let \(\mathcal{B}_l\) be a sub-bundle of \(\mathcal{N}\) such
that \(\forall x \in C_{\max}\) the fiber \((\mathcal{B}_l)_x\) of
\(\mathcal{B}_l\) is a set of all vectors \(v\) of \(\mathcal{N}_x\)
satisfying the following inequality: \(\|v\|< l\) (with respect to the metric
\(g\)).

Let \(\gamma_v(x)(t)\) (\(x \in M,v \in T_x M, t \in \mathbb{R}\)) be a
geodesic line, i.e., the solution of the following ordinary differential
equation
\begin{eqnarray}
\nabla_{\dot{\gamma}_v(x)(t)}\dot{\gamma}_v(x)(t)=0
\end{eqnarray}
with the following initial conditions
\begin{eqnarray}
\gamma_v(x)(0)&=&x,\nonumber\\
\dot{\gamma}_v(x)(t)|_{t=0}&=&v.
\end{eqnarray}
Here \(\nabla_v\) is a Levi-Civita connection on \(M\) with respect to the
metric \(g\). The solution of this differential equation is defined on the
whole real line because \(M\) is compact.

Let \(F_l\) for \(l \in (0,+\infty)\) be a map \(\mathcal{B}_l\rightarrow M\)
which assigns to each point \((x,v) \in \mathcal{B}_l\) (\(x \in C_{\max}, v
\in (\mathcal{B}_l)_x\)) the point \(\gamma_v(x)(1)\). It follows from the
inverse function theorem that there exits a number \(l_0>0\) such that  \(
F_l\) is a diffeomorphism on its image for all \(l:0<l\leq l_0\). $\Box$

\begin{lemma}
If \(\varepsilon\) is small enough then \(\forall \mu:
\mu_{\max}>\mu>\mu_{\max}-\varepsilon\) the set
\(\Gamma(\mu)=f^{-1}(\mu)\) is connected.
\end{lemma}
\textbf{Proof.} Let \(l_0\) be a number from the previous Lemma. It follows
from the Morse Lemma that for every \(x \in C_{\max}\) we can choose
coordinates \(z_1,\dots,z_{d-d_{n}}\) on \((\mathcal{B}_{l_0})_x\) in some
neighborhood \(U\) of zero such that
\begin{eqnarray}
f\circ F_{l_0}|_U=z_1^2+\dots+z_{d-d_{n}}^2.
\end{eqnarray}
Moreover, from construction of these coordinates it follows that in some
neighborhood of every point \(x_0 \in C_{\max}\) they are differentiable
functions of $x$. Therefore, there exists a finite covering
\(\{U_i\}_{i=1,\dots,q}\) of \(C_{\max}\) by open connected sets and a family
of diffeomorphisms \(g_i:U_i\times B_{d-d_n}\rightarrow \pi^{-1}(U_i)\)
\((i=1,\dots,q)\) on its image commuting with the projections such that
\begin{eqnarray}
f\circ F_{l_0}\circ g_i=z_1^2+\dots+z^2_{d-d_{n}},\qquad
i=1,\dots,q.
\end{eqnarray}
Here \(z_i\), \(i=1,\dots,d-d_{n}\) are some coordinates on the ball
\(B_{d-d_n}\) and \(\pi\) is a canonical projection from \(\mathcal{B}_{l_0}\)
to \(C_{\max}\).

We now prove that for every \(l_1:0<l_1<l_0\) there exists \(\varepsilon_1>0\)
such that \(\forall \mu: \mu_{\max}- \varepsilon_1 <\mu\leq \mu_{\max}\),
\(\Gamma(\mu) \subset F_{l_1}\mathcal{B}_{l_1}\). Suppose that \(\forall
n=1,2,\dots\) there exists a point \(x_n\) such that \(f(x_n)>\mu_{\max}-1/n\)
and \(x_n \notin F_{l_1}\mathcal{B}_{l_1}\). Because \(M\setminus
F_{l_1}\mathcal{B}_{l_1}\) is compact, then there exists a point \(x_0 \in
M\setminus F_{l_1}\mathcal{B}_{l1}\) and sub-sequence \(\{x_{n_k}\}\) of
\(\{x_n\}\) such that \(x_{n_k}\rightarrow x_0\) as \(k \rightarrow \infty\).
We find that \(f(x_0)=\mu_{\max}\) and \(x_0 \in C_{\max}\). This
contradiction proves our statement. If \(l_1\) is small enough then
\(\mathcal{B}_{l_1}\cap U_i\subset g_i(U_i\times B_{d-d_n})\) for all
\(i=1,\dots,q\). Therefore if \(\mu>\mu_{\max}-\varepsilon_1\) then
\(f^{-1}(\mu)\cap \pi^{-1}(U_i)\subset g_i(U_i\times B_{d-d_n})\) and
connected. So we find that \(f^{-1}(\mu)\) is connected if
\(\mu>\mu_{\max}-\varepsilon_1\). $\Box$

\begin{lemma}
Suppose that for some \(\mu: \mu_i<\mu<\mu_{i+1}\) \((i=1,\dots,n-1)\) the set
\(\Gamma(\mu)\) is connected. Then \(\forall \mu\) such that
\(\mu_i<\mu<\mu_{i+1}\), the set \(\Gamma(\mu)\) is connected.
\end{lemma}
\textbf{Proof.} Let \(\nu \in \mathbb{R}:\mu_i<\nu<\mu_{i+1}\). Let us prove
that \(\Gamma(\nu)\) is connected. We can assume that \(\nu<\mu\), and let
\(\varepsilon\) be a positive number such that
\(\mu_i<\nu-\varepsilon<\mu+\varepsilon<\mu_{i+1}\). Consider the following
sets
\begin{eqnarray}
U_\varepsilon&=&\{x|\nu-\varepsilon<f(x)<\mu+\varepsilon\},
\nonumber\\
\overline{U}_\varepsilon&=&\{x|\nu-\varepsilon\leq
f(x)\leq\mu+\varepsilon\}
\end{eqnarray}
Consider also the following differential equation on \(M\)
\begin{eqnarray}
\dot{\gamma}(t)=\frac{\rm grad \mit f(\gamma(t))}{\|\rm
grad \mit f(\gamma(t)) \|^2}.\label{1}
\end{eqnarray}
(Recall that \(M\) has a Riemann metric). The right hand side of this equation
is well defined on \(U_\varepsilon\). The solution of (\ref{1})is
$\gamma_x(t)$ with the initial condition
\begin{eqnarray}
\gamma_x(0)=x,\qquad x \in \Gamma(\mu).
\end{eqnarray}
By the extension theorem~\cite{pontriagin,arnold} this solution must leave the
compact set \(\overline{U}_{\varepsilon/2}\). It is easy to prove that
\(f(\gamma_x(t))=t+\mu\). So the solution \(\gamma_x(t)\) is defined and
unique on the interval \((\nu-\mu-\varepsilon/3,\mu+\varepsilon/3)\).
Therefore we have a smooth map
\(\Delta_{\mu,\nu}:\Gamma(\mu)\rightarrow\Gamma(\nu)\), \(x\mapsto
\gamma_x(\nu-\mu)\). By the same means we can construct the map
\(\Delta_{\nu,\mu}:\Gamma(\nu)\rightarrow\Gamma(\mu)\).
\(\Delta_{\mu,\nu}(x)=y\) if and only if \(x\) and \(y\) lie on the same
integral curve of (\ref{1}). We have \(\Delta_{\mu,\nu}\circ
\Delta_{\nu,\mu}={\rm id}\) and \(\Delta_{\nu,\mu}\circ \Delta_{\mu,\nu}={\rm
id}\). So \(\Gamma(\mu)\) and \(\Gamma(\nu)\) are diffeomorphic. $\Box$

\begin{lemma}
Suppose that the assumptions of the theorem hold. Let \(\mu \in \mathbb{R}\):
\(\mu_i<\mu<\mu_{i+1}\), \(\mu_i=2,\dots,n-1\), and \(\Gamma(\mu)\) is
connected. Then \(\forall \nu\) such that \(\mu_{i-1}<\nu<\mu_{i}\), the set
\(\Gamma(\nu)\) is also connected.
\end{lemma}
\textbf{Proof.} We prove this lemma only for the case of connected \(C_i\).
The general case is analogous to this case.

As in Lemma~\ref{m:1}, let \(\mathcal{B}_l\) be a bundle with the base \(C_i\)
which consists of all vectors \(v\) normal to \(C_i\) and such that \(\|v\|<
l\). We have \(\mathcal{B}_{l_1}\subset\mathcal{B}_{l_2}\) for \(l_1<l_2\).
Let \(F_l\) be a map \(\mathcal{B}_l\rightarrow M\) constructed as in Lemma 1.
As in Lemma~\ref{m:1}, we find that \(F_l\) is a diffeomorphism if \(0<l\leq
l_0\) for some positive number \(l_0\). As in Lemma 1 we find that for every
\(l_0'<l_0\) there exists a covering \(\{U_j\}_{j=1,\dots,p}\) of \(C_i\) by
open connected sets and the family of diffeomorphisms \(g_j:U_j\times
B_{d-d_i}\rightarrow \pi^{-1}(U_j)\) on its image commuting with the
projections such that
\begin{eqnarray}
f\circ F_l\circ
g_j=z_1^2+\dots+z^2_{\lambda_i^+}-z^2_{\lambda_i^+
+1}-\dots-z^2_{d-d_i}+\mu_i
\end{eqnarray}
Here \(B_{d-d_i}\) is a \(d-d_i\)-dimensional ball and \(\pi\) a canonical
projection from \(\mathcal{B}_{l_0'}\) to \(C_i\).

It is easy to see that for every \(l_0'<l_0\) there exists a positive number
\(l_1<l_0'\) such that \(\forall j=1,\dots,p\) \(\mathcal{B}_{l_1}\cap
\pi^{-1}(U_j) \subset g_j(U_j\times B_{d-d_i})\). For every \(l_1<l_0'\) there
exists a positive number \(\varepsilon_2\) such that \(\forall x \in C_i\),
\((\mathcal{B}_{l_1/2})_x\cap
F_{l_0}^{-1}(\Gamma(\mu_i+\kappa))\neq\emptyset\) \(\forall\kappa
:|\kappa|<\varepsilon_2\). We now prove that \(\mathcal{B}_{l_1}\cup
\pi^{-1}(U_j)\cap F_{l_0}^{-1}( \Gamma(\mu+\kappa))\) is connected \(\forall
j=1,\dots,p\) if \(|\kappa|<\varepsilon_2\). Indeed, let \(x_1\) and \(x_2\)
be two points which lie in the set \(\mathcal{B}_{l_1}\cup \pi^{-1}(U_j)\cap
F_{l_0}^{-1}(\Gamma(\mu+\kappa))\). We can consider only the case
\(\kappa>0\). The set \(g_j(U_j\times B_{d-d_i})\cap
F^{-1}_{l_0}(\Gamma(\mu_i+\kappa))\) is diffeomorphic to
\(\mathbb{R}^{\lambda_i^+}\times S^{\lambda_i^--1}\times U_i\) and connected.
Let \(\gamma(t)\;t \in [0,1]\) be a path in \(g_j(U_j\times B_{d-d_i})\cap
F^{-1}_{l_0}( \Gamma(\mu_i+\kappa))\) such that \(\gamma(0)=x_1\),
\(\gamma(1)=x_2\). Let \(d(x)\) be a function on \(\mathcal{B}_{l_0}\) defined
as follows: \(d((z,v))=\|v\|^2\), where \(z \in C_i\) and \(v \in
(\mathcal{B}_{l_0})_x\). Let \(x \in F_{l_0}^{-1}(\Gamma(\mu+\kappa))\cap
g_j(U_j\times B_{d-d_{i}})\) and \(w(x)\) be a projection of \(\nabla d(x)\)
to the tangent space of \(F_{l_0}^{-1}(\Gamma(\mu+\kappa))\) at \(x\). It is
obvious that \(w(x)\neq 0\) \(\forall x \in
F_{l_0'}^{-1}(\Gamma(\mu+\kappa))\cap g_j(U_j\times
B_{d-d_{i}})\cap(\mathcal{B}_{l_0'}\setminus \mathcal{B}_{l_1})\) if \(l_0'\)
is a sufficiently small number. So we can retract the path \(\gamma(t)\) along
the vector field \(w\) to the part \(\tilde{\gamma}(t)\) which lies in
\(\mathcal{B}_{l_1}\) and connects the points \(x_1\) and \(x_2\).  So
\(\mathcal{B}_{l_1}\cup \pi^{-1}(U_j)\cap F_{l_0}^{-1}( \Gamma(\mu+\kappa))\)
is connected. Now we can find that \(\mathcal{B}_{l_1}\cup \cap F_{l_0}^{-1}(
\Gamma(\mu+\kappa))\) is connected.

Now let \(x_1, x_2 \in \Gamma(\mu)\), \(\mu<\mu_i\),
\(|\mu-\mu_i|<\varepsilon_3\).  Let \(U=F_{l_0}(\mathcal{B})_{l_1/2}\),
\(V=F_{l_0}(\mathcal{B})_{l_1/3}\), \(W=F_{l_0}(\mathcal{B})_{l_1/4}\). At
first suppose that \(x_1 \notin U\) and \(x_2 \notin U\). Let
\(\gamma_{x_1}(t)\), \(\gamma_{x_2}(t)\) be solutions of the differential
equation (\ref{1}) with initial conditions \(x_1\) and \(x_2\) respectively.
The paths \(\gamma_{x_1}(t)\) and \(\gamma_{x_2}(t)\) intersect the
sub-manifold \(\Gamma(\mu+\varepsilon_3)\) at the points \(y_1\) and \(y_2\)
if \(\varepsilon_3\) is enough small. Let \(\widetilde{\delta}(t), t \in
[0,1]\) be a path such that \(\forall t \in [0,1]\; \widetilde{\delta}(t) \in
\Gamma(\mu+\varepsilon)\) and \(y_1=\widetilde{\delta}(0),
y_2=\widetilde{\delta}(1)\). We must consider the following two cases.

1) \(\widetilde{\delta}\cap V=\emptyset\). If \(\varepsilon_3\) is small
enough then we can deform the part \(\widetilde{\delta}\) along the vector
field \(\nabla {f}/\| \nabla f\|^2\) to the part \(\delta\) which lies on
\(\Gamma(\mu)\) and connects the points \(x_1\) and \(x_2\).

2) \(\widetilde{\delta}\cap V \neq\emptyset\). If
\(\varepsilon_3\) is small enough then \(y_1,y_2 \notin
V\). We can decompose the part \(\widetilde{\delta}\) as
\(\widetilde{\delta}=\widetilde{\alpha}_1\circ\widetilde{\beta}\circ\tilde{\alpha}_2\),
where
\begin{eqnarray}
\widetilde{\alpha_2}(1) \in \partial V,\qquad\forall t \in
[0,1]\; \widetilde{\alpha_2}(t)
\notin V \nonumber\\
\widetilde{\alpha_1}(0) \in \partial V,\qquad\forall t \in
[0,1]\; \widetilde{\alpha_1}(t) \notin V.
\end{eqnarray}
If \(\varepsilon_3\) is a sufficiently small positive number we can deform the
paths \(\widetilde{\alpha}_1\) and \(\widetilde{\alpha}_2\) along the vector
field \(\nabla f/\|\nabla f\|^2\) into the the paths \(\alpha_1,\alpha_2
\subset \Gamma(\mu)\) such that \(\alpha_1, \alpha_2 \nsubseteq W\) and
\(\alpha_2(0)=x_1, \alpha_2(1)\in U\), \(\alpha_1(1)=x_2\) and
\(\alpha_1(0)\in U\). But, it has been proved that \(U\cap\Gamma(\mu)\) is
connected. Therefore, there exists a path \(\beta \subset \Gamma(\mu)\) such
that \(\beta(1)=\alpha_1(0)\) and \(\alpha_2(1)=\beta(0)\). We see that the
path \(\alpha_1\circ\beta\circ\alpha_2\) connects the point \(x_1\) and
\(x_2\).

Consideration of the case with \(x_1 \in U\) or \(x_2 \in \mathcal{B}_U\) is
analogous to consideration of the previous case. The statement of the theorem
follows from these four Lemmas. $\Box$

\begin{theorem}
Each level set of the objective function $J[\cdot,\bw]$
defined by~(\ref{eq7}) is connected.
\end{theorem}
{\bf Proof.} The objective function $J[\cdot,\bw]$ is a generalized Morse
function. The sub-manifold of solutions corresponding to the global maximum in
the coordinates $\tilde u_1,\tilde u_2\in\mathbb C^4$ is defined by $\|\tilde
u_1\|=\|\tilde u_2\|=1$, $\langle \tilde u_1,\tilde u_2\rangle=0$. It is a
Stiefel manifold, ${\cal M}^\bw_{\rm max}=V_2(\mathbb C^4)$, and hence is
connected. The Morse indices of the function $J[\cdot,\bw]$ are $\nu_\pm>2$ at
any saddle sub-manifold. Therefore this function satisfies the conditions of
Theorem~\ref{t1} and its each level set is connected. $\Box$

\section{Conclusions}
In this paper the landscape of the objective functions for open quantum
systems controlled by general Kraus maps is investigated in detail for the
two-level case. It is shown that a typical objective function has: (a) no
false traps, (b) multi-dimensional sub-manifolds of the optimal global
solutions, and (c) each level set is connected. These results may be
generalized to systems of arbitrary dimension $N$, although a full enumeration
of the critical sub-manifold dimensions remains open for analysis. The
landscape analysis and the conclusions rest on assuming that the controls can
manage the system and the environment. Managing the environment, in practice,
is likely not highly demanding, as control over only the immediate environment
of the system is most likely needed. The critical point topology of general
controlled open system dynamics could provide a basis to explain the relative
ease of practical searches for optimal solutions in the laboratory, even in
the presence of an environment.

\section*{Acknowledgments}This work was supported by the Department of Energy. A. Pechen
acknowledges partial support from the grant RFFI 05-01-00884-a. The authors
thank Jonathan Roslund for help with drawing the Figure~\ref{fig1}.

\end{document}